\documentclass[preprint,review,12pt]{elsarticle}

\usepackage[utf8]{inputenc}
\usepackage{amssymb}
\usepackage{picins}
\usepackage[]{graphicx}
\usepackage{setspace}
\usepackage{tocloft}
\usepackage{amsmath}
\usepackage{amssymb}
\usepackage{caption}
\usepackage{mwe}
\usepackage{graphbox}
\usepackage{subcaption}
\usepackage{algpseudocode}
\usepackage[ruled,linesnumbered,vlined]{algorithm2e}
\usepackage{color, colortbl}
\usepackage{array}
\usepackage{hyperref}
\usepackage{multirow}
\usepackage{multicol}
\usepackage{float}
\usepackage[capitalise]{cleveref}
\usepackage{textcomp}
\usepackage{changepage}
\graphicspath{{pics/}{fig/}}

\journal{Computerized Medical Imaging and Graphics}

\begin{document}

\begin{frontmatter}



\title{QC-SPHRAM: Quasi-conformal Spherical Harmonics Based Geometric Distortions on Hippocampal Surfaces for Early Detection of the Alzheimer's Disease}


\author[1]{Anthony Hei-Long CHAN}
\ead{hlchan@math.cuhk.edu.hk}

\author[2]{Yishan LUO}
\ead{lys84818@hotmail.com}

\author[2]{Lin SHI}
\ead{shilin@cuhk.edu.hk}

\author[3]{Lok Ming LUI\corref{mycorrespondingauthor}}
\cortext[mycorrespondingauthor]{Corresponding author}
\ead{lmlui@math.cuhk.edu.hk}

\address[1]{222B, Lady Shaw Building, The Chinese University of Hong Kong, Shatin, Hong Kong }
\address[2]{BrainNow Medical Technology Limited, Unit 201, 2/F., Lakeside 2, No. 10 Science Park West Avenue, Hong Kong Science Park, Pak Shek Kok, N.T., Hong Kong}
\address[3]{207, Lady Shaw Building, The Chinese University of Hong Kong, Shatin, Hong Kong }

\begin{abstract}
This paper examines the importance of surface geometry of hippocampus for the analysis of the Alzheimer's disease (AD). We propose a disease classification model, called the QC-SPHARM, for the early detection of AD. The proposed QC-SPHARM can distinguish between normal control (NC) subjects and AD patients, as well as between amnestic mild cognitive impairment (aMCI) patients having high possibility progressing into AD and those who do not. Using the spherical harmonics (SPHARM) based registration, hippocampal surfaces segmented from the ADNI data are individually registered to a template surface constructed from the NC subjects using SPHARM. Local geometric distortions of the deformation from the template surface to each subject are quantified in terms of conformality distortions and curvatures distortions. The measurements are combined with the spherical harmonics coefficients and the total volume change of the subject from the template. Afterwards, a t-test based feature selection method incorporating the bagging strategy is applied to extract those local regions having high discriminating power of the two classes. The disease diagnosis machine can therefore be built using the data under the Support Vector Machine (SVM) setting. Using 110 NC subjects and 110 AD patients from the ADNI database, the proposed algorithm achieves $85.2\%$ testing accuracy on 80 random samples as testing subjects, with the incorporation of surface geometry in the classification machine. Using 20 aMCI patients who has advanced to AD during a two-year period and another 20 aMCI patients who remain non-AD for the next two years, the algorithm achieves $81.2\%$ accuracy using 10 randomly picked subjects as testing data. Our proposed method is $6\%-15\%$ better than other classification models without the incorporation of surface geometry. The results demonstrate the advantages of using local geometric distortions as the discriminating criterion for early AD diagnosis.
\end{abstract}

\begin{keyword}
Hippocampus\sep Alzheimer's Disease\sep Amnestic Mild Cognitive Impairment\sep Disease Classification\sep Medical Image Analysis


\end{keyword}

\end{frontmatter}


\section{Introduction}

Recent research show that the Alzheimer's disease (AD) population is growing rapidly due to the world-wide ageing problem \cite{AD_population_1,AD_population_2,AD_population_3,AD_population_4}. As an incurable disease, early detection is particularly crucial in the perspective of early intervention to slow down the progress of the disease. However, definite clinical diagnosis of AD in prodromal stage is merely impossible as significant symptoms in terms of memory impairment only shows in a late stage of the disease. More often, patients suffering from the prodromal stage of AD are clinically classified as amnestic mild cognitive impairment (aMCI). And the discovery that not all aMCI patients would progress to AD adds one another difficulty to an early diagnosis. 

Fortunately, medical research proves that the brain structure would undergo specific and measurable deformation since early AD. As a result, analyzing the MR brain images provides a possible solution to a trustworthy early diagnosis. Over years, medical research has been carried out and different criteria to analyze the brain for the AD detection has been proposed. In particular, many reports conclude that the hippocampus (Hipp) is affected in the earliest stage of AD \cite{AD_Hipp_1,AD_Hipp_2,AD_Hipp_3}. Since then, analyzing the hippocampal atrophy for early AD detection is deeply studied. Most research takes the relative volume change of Hipp as the main criteria for classifying AD from aMCI and normal control (NC) subjects, as supported by medical reports that Hipp atrophy is a common sign in AD progression \cite{Hipp_volume_1,Hipp_volume_2,Hipp_volume_3,Hipp_volume_4,Hipp_volume_5,Hipp_volume_6}. The discriminative power of Hipp volume in AD detection might be improved after considering more features on the Hipp surface \cite{AD_Hipp_compare,AD_CNN}.

The difficulty in utilizing other Hipp surface features lie on the fact that there is no recognizable biomarker on the Hipp surface \cite{Hipp_landmark}. It is thus hard to obtain a reliable surface correspondence. Nevertheless, Hipp is known to have a genus-0 topology, i.e. an object without holes. As such, by the theory of partial differential equation, the geometry of Hipp can be described by a sequence of functions called the spherical harmonics (SPHARM) \cite{spharm}. Each member of the sequence describes certain global geometry of the corresponding Hipp. The correspondence between an Hipp an its corresponding sequence of SPHARM coefficients is unique up to an isometry. Therefore, utilizing the SPHARM coefficients for AD diagnosis has also be studied by a number of researchers \cite{Hipp_spharm_1,Hipp_spharm_2,Hipp_spharm_3,Hipp_spharm_4}. However, such approach fails to report an expectedly high classification accuracy. Not only numerical errors may occur due to bad meshing of the Hipp surface, the need to truncate the SPHARM sequence to lower order terms so as to improve the computational cost greatly limits the discrimative power of the method. Therefore, a better approach over the volumetric method and the SPHARM-based method is still in demand. 

In this work, we propose the QC-SPHARM model to utilize the Hipp surface geometric distortion combined with other metrics including volume changes for early AD detection. Given a database of pre-labelled brain scan (.nii extension), we use the AccuBrain\textsuperscript{\tiny\textregistered} \cite{AccuBrain} algorithm to segment the left Hipp volume from the data. The process is validated and edited by neuroscientists to ensure the segmentation accuracy. We then use the ITK-SNAP\textsuperscript{\tiny\textregistered} \cite{itk-snap} to extract the surface from the volume. To compare surfaces based on their surface geometry, a pertinent mean template surface plays an important role. In this work, we apply the SHREC scheme \cite{SHREC} (which is based on the SPHARM theory) to construct a mean template surface of all NC subjects. The mean surface demonstrates the overall geometry a normal Hipp. Each Hipp is then registered to this mean surface using the SPHARM registration. The truncated SPHARM coefficients of each Hipp is imposed onto the mean surface accordingly. This results in a vertex-wise correspondence between each Hipp and hence the local geometric distortions of each Hipp from the mean surface at each vertex is studied. We measure the local geometric distortions in terms of conformality distortions and curvatures distortions. Combining all these measurements result in a shape index which gives a complete description of the geometric distortions involved. That is, if the shape index of two Hipps are identical, then the two Hps are also identical up to a rigid motion. The shape index is then combined with the SPHARM and/or the volume distortion of the corresponding Hipp from the mean surface to form a feature vector. After applying a bagging strategy incorporating t-test algorithm to further enhance the discriminative power of it, the support vector machine (SVM) is used to build a machine to classify AD subjects from the others. Whenever a new (unlabelled) subject is present, by registering his segmented Hipp surface to the mean surface and then computing the shape index of it, the SVM machine can automatically label if the subject belongs to the AD class. The proposed framework is partly inspired by \cite{qcpaper_1}.

The proposed QC-SPHARM model has three main contributions. Firstly, the automatic generation of the template reference surface allows the study of surface geometric deformations without requesting longitudinal data (i.e. different time frame) from the same patient. After being scanned once, the Hipp surface is registered to the mean surface and this deformation provides much important measurements to the classification machine. This allows rapid diagnosis and facilitates the medication pipeline a lot. Secondly, the QC-SPHARM model combines the local geometric distortions, including the conformality distortion, the Gaussian curvature distortion and the mean curvature distortion, with the SPHARM coefficients and the global volume distortion to form the feature vector. Such joint-force of different measurements provides more freedom on the choices of the discriminating features and improves the classification accuracy. Lastly, the QC-SPHARM model applies the SVM on the feature vectors mentioned above, which involves both local and global geometric measurements. With the Gaussian radial basis function (RBF) kernel used, the classification model is both accurate and stable.

The proposed framework is tested with subjects from the Alzheimer's Disease Neuroimaging Initiative (ADNI) database \footnote{Data used in preparation of this article were obtained from the Alzheimer’s Disease Neuroimaging Initiative
(ADNI) database (adni.loni.usc.edu). As such, the investigators within the ADNI contributed to the design and implementation of ADNI and/or provided data but did not participate in analysis or writing of this report. A complete listing of ADNI investigators can be found at:
\url{http://adni.loni.usc.edu/wp-content/uploads/how_to_apply/ADNI_Acknowledgement_List.pdf}
} and the results are compared with other traditional models on the same dataset. It is evident that the proposed framework is accurate in classifying AD from normal control subjects, as well as detecting aMCI patients having a higher possibility of progressing into AD.

\section{Data}

There are two main aspects in AD diagnosis. On one hand, it is important to be able to detect AD subjects from those NC subjects. To validate the accuracy of the proposed model in this aspect, we collected a dataset consisting of 110 AD patients and 110 NC subjects from the ADNI database. On the other hand, it is reported that aMCI is an intermediate stage in the progression of AD. Whether an aMCI patient would progress into AD is hence crucial for early AD diagnosis. The MCI-AD converter is defined when a subject’s diagnostic status has advanced during the two-year period. We use the aMCI-AD label to denote subjects who transitioned from MCI to AD from baseline to 24-month follow-up examinations. And we also use the aMCI-stable label to denote those who kept his/her baseline diagnosis for the whole two-year period. In this work, we collected a dataset consisting of 20 aMCI-stable patients and 20 aMCI-AD patients. The MRI data analyzed in this paper are all baseline scans. Table (\ref{tab:database}) summarizes the composition of the two database involved in this work.
\begin{table}[h]
\makebox[1 \textwidth][c]{\resizebox{1.05 \textwidth}{!}{
    \centering
    \begin{tabular}{|c|c|c|c|c|c|}
    \hline
    Database & $\#$ Class NC & $\#$ Class AD & $\#$ Class aMCI-stable & $\#$ Class aMCI-AD\\
    \hline
    A & 110 & 110 & - & - \\
    B & - & - & 20 & 20 \\
    \hline
    \end{tabular}}}
    \caption{Composition of the 2 datasets involved}
    \label{tab:database}
\end{table}

\begin{figure}[h]
    \centering
    \includegraphics[width=.8\textwidth]{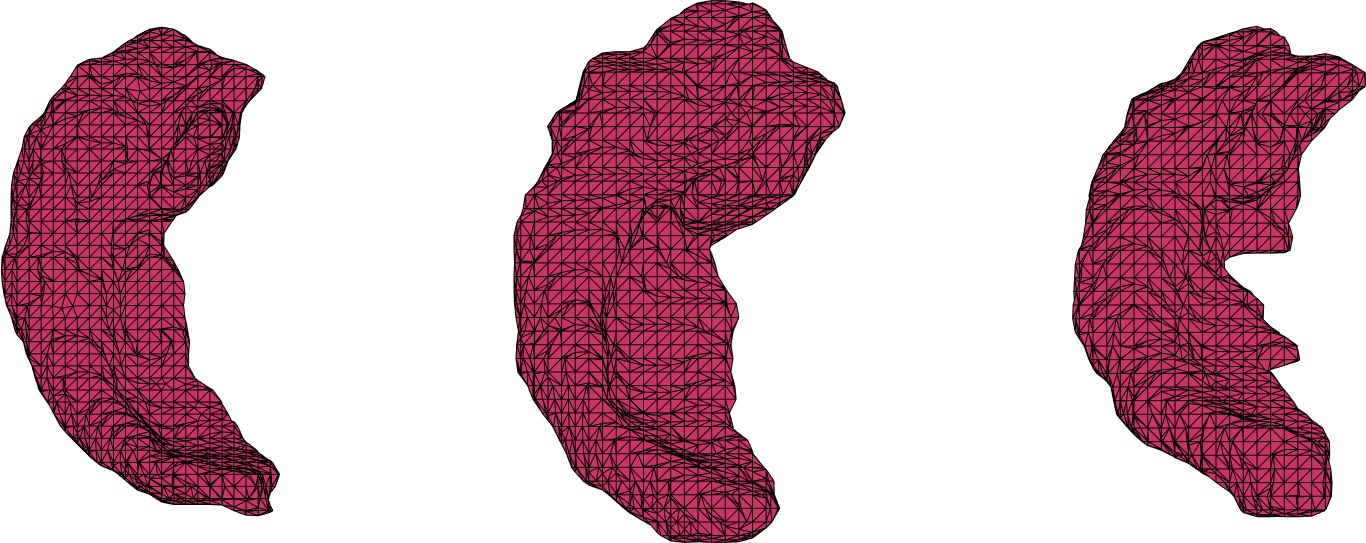}
    \caption{Examples of segmented hippocampal surfaces in the database}
    \label{fig:hipp_mesh}
\end{figure}

The left hippocampus (Hipp) of each subject in the ADNI database is segmented using the AccuBrain\textsuperscript{\tiny\textregistered} \cite{AccuBrain}. The ITK-SNAP\textsuperscript{\tiny\textregistered} \cite{itk-snap} is then used to further extract the surface from the Hipp volume data. \cref{fig:hipp_mesh} shows some examples of the segmented left hippocampal surface. Each surface consists of approximately 2,500 vertices.

\section{Method}

The proposed QC-SPHARM model is explained in this section in detail. The SPHARM-based registration method is explained in the first subsection. After obtaining a vertex-wise correspondence between different Hipps, the local geometric distortions from the mean surface can be computed, which is the main scope of the second subsection. Finally, the geometric measurements can be combined with other quantities such as the SPHARM coefficients and the relative Hipp volume to form a feature vector. For easier discussion, we elaborate the following discussions in reference to the dataset A, i.e. NC versus AD patients. Most steps can be followed directly by replacing NC subjects by aMCI-stable patients, and AD patients by aMCI-AD patients respectively.

Data used in the preparation of this article were obtained from the Alzheimer’s Disease Neuroimaging Initiative (ADNI) database (adni.loni.usc.edu). The ADNI was launched in 2003 as a public-private partnership, led by Principal Investigator Michael W. Weiner, MD. The primary goal of ADNI has been to test whether serial magnetic resonance imaging (MRI), positron emission tomography (PET), other
biological markers, and clinical and neuropsychological assessment can be combined to measure the progression of mild cognitive impairment (MCI) and early Alzheimer’s disease (AD)

\subsection{Surface registration}
\label{subsec:surface_reconstruction}

In this work, we propose to detect AD by analyzing the geometric deformation of each Hipp surface to a mean surface. The construction of a pertinent mean surface is crucial.

\begin{figure}[h]
    \centering
    \includegraphics[width=.8\textwidth]{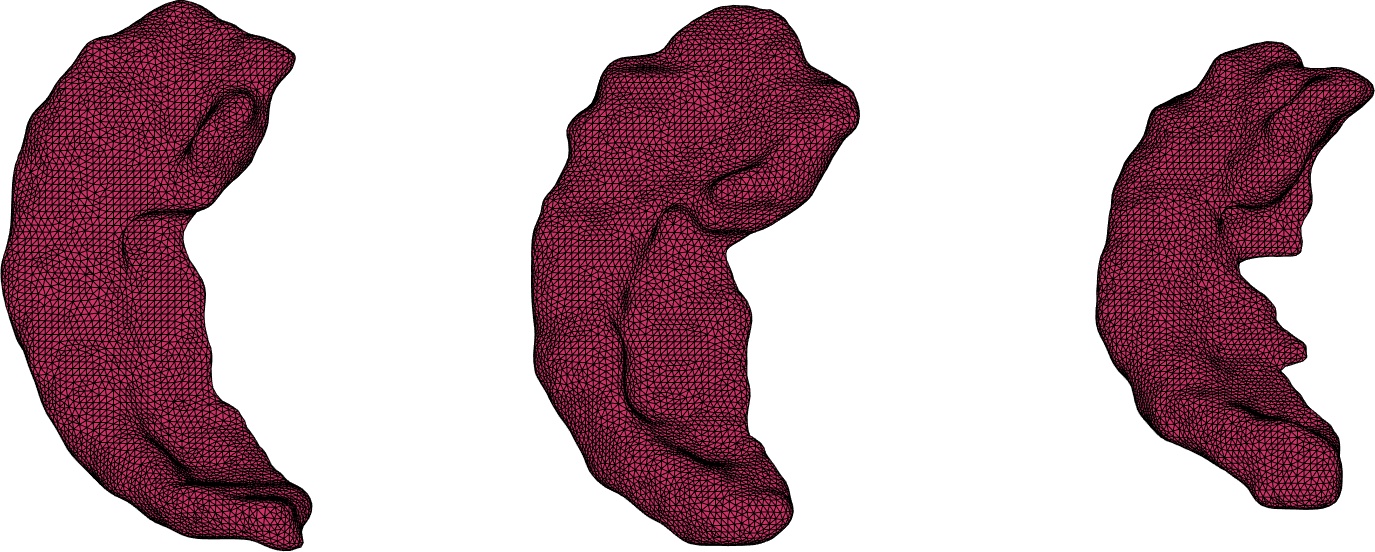}
    \caption{Examples of improved hippocampal surfaces}
    \label{fig:hipp_mesh_improved}
\end{figure}

In our QC-SPHARM model, the first task is to construct a relevant mean surface and elaborate a mutual vertex-wise correspondences between subjects to facilitate comparison of local geometric distortions. In prior to this step, we need to improve the mesh quality in terms of triangulation regularity since it is a crucial factor to the accuracy of the registration. In this work, the mesh quality is improved by applying the Laplacian smoothing to each subject, followed by a mesh simplification step and then a a mesh refinement step to improve the vertex density of the surfaces. The edge lengths of the resultant meshes are less deviated. \cref{fig:hipp_mesh_improved} demonstrates the sample surfaces as in \cref{fig:hipp_mesh} after the quality improvement process.

Denote each Hipp surface (after quality improvement) by $\hat{M_i}=(\hat{V}_i,\hat{F_i})$, where $\hat{V_i}$ and $\hat{F_i}$ is the vertex set and the face set of $\hat{M_i}$ respectively. To register between surfaces, we use the spherical harmonics (SPHARM) to establish fully automatic registration. Note that each Hipp surface must have a spherical topology. In the SPHARM theory \cite{spharm_morphing}, there is a sequence of Fourier basis function
\begin{equation}
    Y^m_l(\theta,\phi)=\sqrt{\frac{(2l+1)(l-m)!}{4(l+m)!}}P^m_l(\text{cos}\theta)\text{exp}(im\phi),
\label{spharm_basis}
\end{equation}
named the spherical harmonics, that can completely describe any spherical object. In other words, each Hipp surface $\hat{M_i}$ can be represented by
\begin{equation}
    \hat{\mathbf{v}}_i^j=\hat{\mathbf{v}}_i(\theta^j_i,\phi^j_i)=\sum^\infty_{l=0}\sum^l_{m=-l}\mathbf{r}^m_{l,i} Y^m_l(\theta^j_i,\phi^j_i),
\label{spharm}
\end{equation}
for each $\hat{\mathbf{v}}^j_i\in\hat{V_i}$, where $\mathbf{r}^m_{l,i}$ is the 3-dimensional Fourier coefficients (called the SPHARM coefficients) of $\hat{M_i}$ corresponding to $Y^m_l$. The correspondence between each $\hat{M_i}$ and its spherical harmonics is unique. Each term in the RHS of equation (\ref{spharm}) describes part of the geometry of the Hipp surface. It is noted that lower order terms describe more general geometry, and higher order terms describe some finer details of the surface geometry. While including higher order terms of $\{Y^m_l\}$ can increase the discrimative power of the model, in practice it puts large burdens to the computational cost. Therefore, it is common to truncate the high order terms of $\{Y^m_l\}$. In particular, all the terms beyond order $31$ are truncated in this work. Therefore, each $\hat{M_i}$ is approximated by
\begin{equation}
    \hat{\mathbf{v}}_i(\theta^j_i,\phi^j_i)\approx\sum^{31}_{l=0}\sum^l_{m=-l}\mathbf{r}^m_{l,i} Y^m_l(\theta^j_i,\phi^j_i).
\end{equation}

To compare the geometry among Hipp surfaces in the database, a reference template surface is needed. In this work, the reference surface is chosen to be the mean surface $M_0$ of all the NC subjects, which is constructed as follows.

\begin{figure}[h]
    \centering
    \includegraphics[width=.5\textwidth]{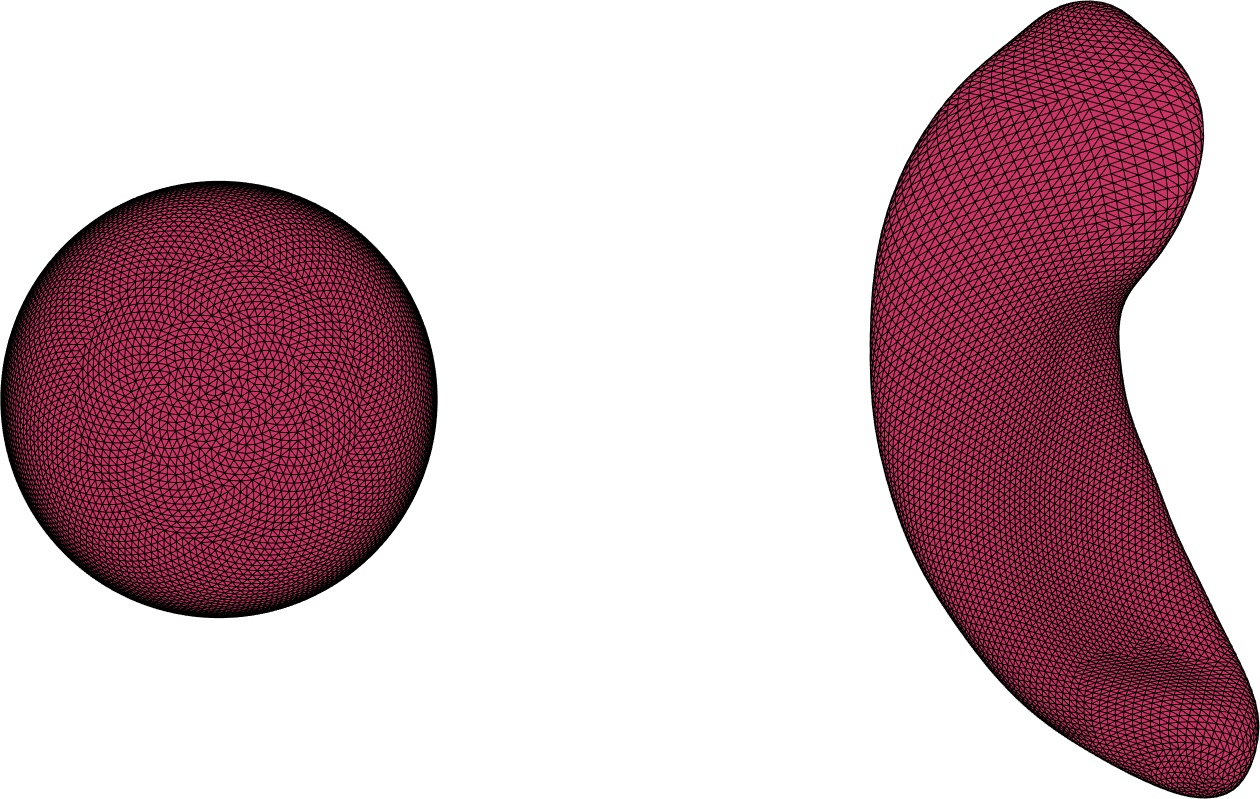}
    \caption{(Left) Spherical template surface $S_0$; (Right) The constructed mean surfaces $M_0$}
    \label{fig:hipp_mean}
\end{figure}

Firstly, we apply the SHREC algorithm \cite{SHREC} on the collection of those NC subjects. The SHREC algorithm mutually aligns the underlying SPHARM parametrization $\{\mathbf{r}^m_{l,i}\}$ of each HP surface to give the parametrization $\{\bar{\mathbf{r}}^m_l\}$ of the mean surface. For more details of the SHREC algorithm, readers are referred to \cite{SHREC}. After constructing the mean parametrization $\{\bar{\mathbf{r}}^m_l\}$, a template spherical mesh $S_0=(\hat{V}_0,F_0)$ is constructed to provide a triangulation to the mean surface. Hence, the mean surface $M_0=(V_0,F_0)$ is constructed by imposing $\{\bar{\mathbf{r}}^m_l\}$ onto $S_0$:
\begin{equation}
    \mathbf{v}^j_0=\mathbf{v}_0(\theta^j,\phi^j)=\sum^{31}_{l=0}\sum^l_{m=-l}\bar{\mathbf{r}}^m_lY^m_l(\theta^j,\phi^j)
\label{mean_surface}
\end{equation}
for each $\mathbf{v}^j_0\in V_0$. In this work, the template mesh is constructed by $N=8,000$ vertices. \cref{fig:hipp_mean} shows the spherical template mesh $S_0$ and the constructed mean surface $M_0$ for reference. It is noted that the meaning of creating a mean template surface is to demonstrate a normal Hipp surface of a healthy subject. Therefore, in manipulating with the dataset B, we still use the above mean surface (\ref{mean_surface}) as the template reference surface in this step.

As for the mutual registration between surfaces, each registered surface $M_i=(V_i,F_i)$ is constructed by imposing the SPHARM coefficients $\{\mathbf{r}^m_{l,i}\}$ of $\hat{M_i}$ onto the spherical meshes $S_0$. That is,
\begin{equation}
    \mathbf{v}^j_i=\mathbf{v}_i(\theta^j,\phi^j)=\sum^{31}_{l=0}\sum^l_{m=-l}\mathbf{r}^m_{l,i}Y^m_l(\theta^j,\phi^j).
\label{mesh_reg}
\end{equation}

\begin{figure}[h]
    \centering
    \includegraphics[width=.8\textwidth]{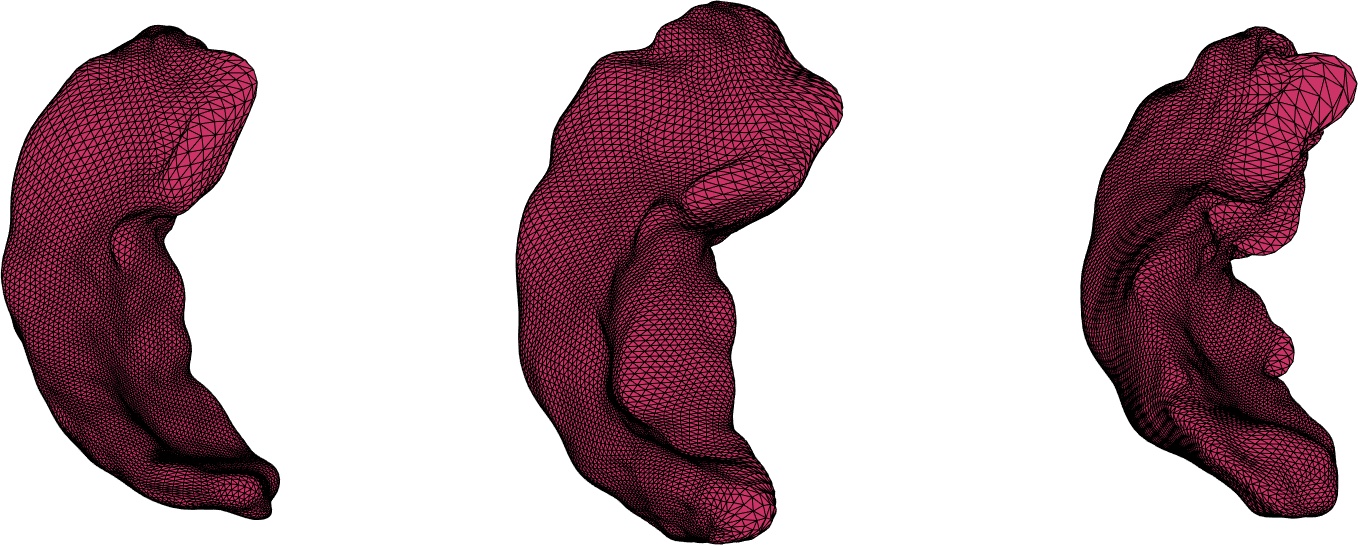}
    \caption{Examples of registered hippocampal surfaces}
    \label{fig:hipp_reg}
\end{figure}

\begin{figure}[h]
    \centering
    \includegraphics[width=.8\textwidth]{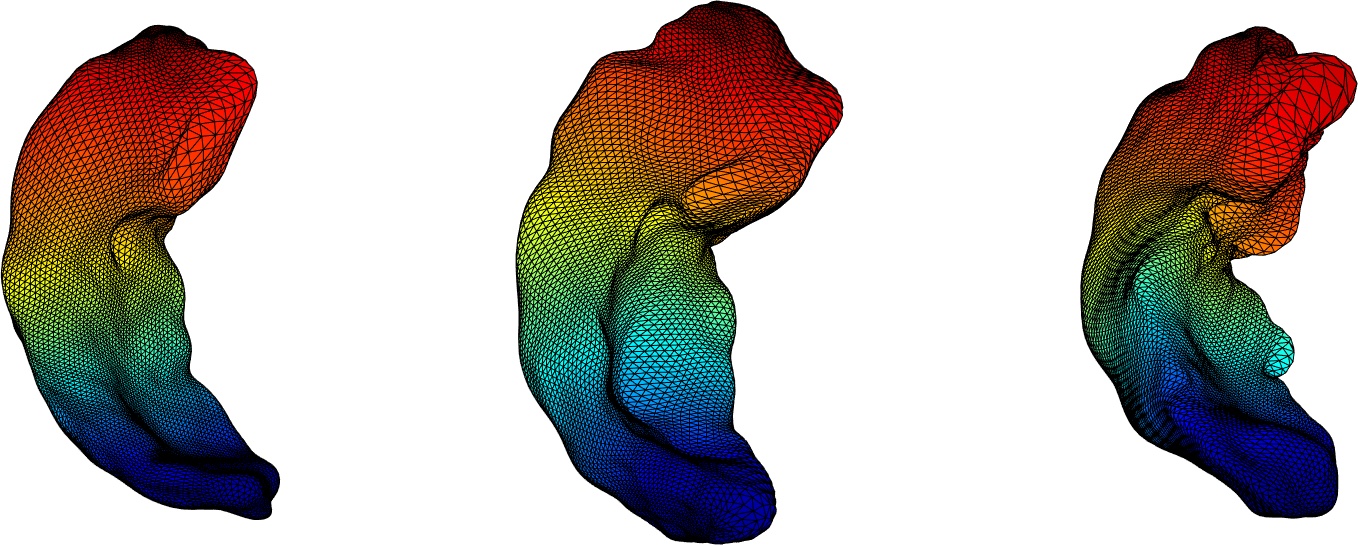}
    \caption{Visualization of the vertex-correspondences of the registered hippocampal surfaces}
    \label{fig:hipp_reg_color}
\end{figure}

Therefore, each surface $M_i$ has the same number of vertices (as that of the mean surface $M_0$) and each vertex $\mathbf{v_i}\in V_i$ is one-to-one corresponded across subject indices $i$. The face set $F_i$ for each subject is indeed inherited from $S_0$, i.e. $F_i=F_0$ for all subjects. \cref{fig:hipp_reg} demonstrates some examples of the registered Hipp surfaces. \cref{fig:hipp_reg_color} further visualizes, by coloring, the vertex-wise correspondences among the surfaces.


\subsection{Geometric measurements}

From the registration step, the surface registration $f_i:M_0\to M_i$ is obtained for each hippocampal surface $M_i$. That is, $f_i(\mathbf{v}^j_0)=\mathbf{v}^j_i$ for all $i,j$. To quantitatively measure the shape deformation, a vertex-wise shape index is designed using the conformality distortion and the curvatures distortions to analyze the geometric distortion of $M_0$ from $M_i$. The shape index $E_{shape}^i: f_i(M_0)\to \mathbb{R}^+$ is defined by
\begin{equation}
E_{shape}^i(\mathbf{v}^j) = \gamma |\mu(f_i)(\mathbf{v}_0^j)| +\alpha |H(\mathbf{v}_0^j) - H(f_i(\mathbf{v}_0^j))| + \beta|K(\mathbf{v}_0^j) - K(f_i(\mathbf{v}_0^j))|,
\label{shape_index}
\end{equation}
\noindent where $\alpha,\beta,\gamma>0$ are weighting parameters. $|\mu(f_i)(\mathbf{v}_0^j)|$ is the conformality distortion term defined by
\begin{equation}
|\mu(f_i)(\mathbf{v}_0^j)| = |\frac{\partial f_i}{\partial \bar{z}}(\mathbf{v}_0^j)|/|\frac{\partial f_i}{\partial z}(\mathbf{v}_0^j)|,
\end{equation}
and $H$ and $K$ are the mean curvature and the Gaussian curvature respectively. In particular, the conformality distortion term $\mu$ detects the local geometric distortion in an infinitesimal scale. Examples of using the conformality distortion to aid detecting abnormal deformations can be found in \cite{qcpaper_1,qcpaper_2}. The shape index is a complete descriptor to measure the deformation of the Hipp surface. If $E^i_{shape}=E^j_{shape}$ for any given $\alpha,\beta,\gamma$, then the corresponding surfaces $M_i$ and $M_j$ are identical up to a rigid motion.

In the proposed QC-SPHARM model, we combine the shape index (\ref{shape_index}) with the SPHARM coefficients and the global volume distortion, such that the feature vector $\mathbf{c}_i$ of each subject $M_i$ reads
\begin{align}
\mathbf{c}_i& = (e_{i,1},e_{i,2},\dots,e_{i,N}|r_{i,1},r_{i,2},\dots,r_{i,K}|v_i),
\label{featurevector_composed}
\end{align}
where $e_{ij} =e_{i,j}= E_{shape}^i(\mathbf{v}^j)$, $r_{i,k}=r_{ik}$ is the collection of all SPHARM coefficients of $M_i$ in sequence, and $v_i$ is the global volume distortion of $M_i$ from the mean surface. Combining all the feature vectors gives the feature matrix $C$:
\begin{equation}
C = \left( \begin{array}{ccc|ccc|c}
e_{1,1} & \ldots & e_{1,N} & r_{1,1}  & \ldots & r_{1,K} & v_1\\
\vdots &     & \vdots & \vdots &  & \vdots &\vdots \\
e_{i,1} & \ldots & e_{i,N} & r_{i,1} & \ldots & r_{i,K} & v_i\\
\vdots &    & \vdots & \vdots &  & \vdots & \vdots \\
e_{M,1} & \ldots & e_{M,N} & r_{M,1} & \ldots & r_{M,K} & v_M \end{array} \right)
\label{featurematrix_composed}
\end{equation}

Each row of $C$ captures the degree of geometric distortions of $M_0$ from $M_i$ at all vertices. Each column of $C$ captures the degree of the geometric distortions at a common vertex of all subjects. 

However, it is noted that in practice, not every observation has the same determining power in classifying the NC/AD class. Including observations having low discriminative power may hinder the classification accuracy. To solve this issue, the QC-SPHARM is incorporated with the t-test to extract a subset $\Omega$ of the original set of all observations, at which the geometric distortions due to the progression of AD is the most obvious. 

To improve the stability of the t-test, the bagging strategy \cite{bagging} is incorporated. Firstly, we rewrite the matrix $C$ in the equation (\ref{featurematrix_composed}) by
\begin{equation}
C = \left( \begin{array}{ccccc}
c_{1,1} & \ldots & c_{1,j} & \ldots & c_{1,N+K+1} \\
\vdots &     & \vdots & & \vdots \\
c_{i,1} & \ldots & c_{i,j} & \ldots & c_{i,N+K+1} \\
\vdots &    & \vdots & & \vdots\\
c_{M,1} & \ldots & c_{M,j} & \ldots & c_{M,N+K+1} \end{array} \right)
\label{featurematrix_renamed}
\end{equation}
We perform $M$ independent iterations. In the $i$-th iteration, the $i$-th subject (i.e. the $i$-th row of $C$ in equation (\ref{featurematrix_renamed})) is excluded. The classical t-test is performed on the remaining $M-1$ subjects to obtain a p-value $p^i_j$ for each $j$-th feature. The $M$ testes are combined by setting
\begin{equation}
    p_j = \min_i(p^i_j).
\label{t_test}
\end{equation}
The statistically significant regions can thus be extracted by the index set:
\begin{equation}
\Omega = \{j_1,j_2,\dots,j_n\}
\label{significant}
\end{equation}
\noindent where the $j_k$'s are selected by the criterion
\begin{equation}
p_{j_k}\leq p_{cut}
\end{equation}
in which $p_{cut}\in(0,1)$ is a global constant threshold parameter. In other words, the QC-SPHARM model extracts all the features having p-value less than or equal to $p_{cut}$.

\subsection{The classification model}

Once the statistical significant region $\Omega$ is extracted, the shape deformations at those statistical significant features can be analyzed and the classification model can be built. For each subject $M_i$, the discriminating feature vector $\widetilde{\mathbf{c}}_i$ can be extracted by
\begin{align}
\widetilde{\mathbf{c}}_i = (c_{i,j_1},...,c_{i,j_k},...,c_{i,j_n}).
\end{align}

Combining the discriminating feature vectors of all subjects together gives the discriminating feature matrix $\widetilde{C}$:
\begin{equation}
\widetilde{C} = \left( \begin{array}{cccccc}
c_{1,j_1} & c_{1,j_2} & \ldots & c_{1,j_k} & \ldots & c_{1,j_n} \\
\vdots & \vdots &   & \vdots  &   & \vdots  \\
c_{i,j_1} & c_{i,j_2} & \ldots & c_{i,j_k} & \ldots & c_{i,j_n} \\
\vdots & \vdots &   & \vdots  &   & \vdots  \\
c_{M,j_1} & c_{M,j_2} & \ldots & c_{M,j_k} & \ldots & c_{M,j_n}\end{array} \right).
\end{equation}
The discriminative feature matrix captures those features providing the most significant information for the desired classification. In the proposed QC-SPHARM model, the SVM is employed to fit the data $\widetilde{C}$. The SVM method finds the best hyperplane $\mathbf{w}\cdot\mathbf{x}-b=0$ to separate the two classes. In practice, the algorithm solves the minimization problem:
\begin{align}
    & \min_{\mathbf{w},b}\frac{1}{2}\mathbf{w}^T\cdot\mathbf{w}\\
    \text{subject to}\quad & y^i\left(\mathbf{w}\cdot \widetilde{\mathbf{c}}^i+b\right)\geq 1
\label{svm}
\end{align}
where $y^i$ is the class label for the $i$-th subject. In this work, we set  $y^i=1$ for NC subjects and $y^i=-1$ for AD subject. To better suit a nonlinear classification environment, the kernel trick policy is used to replace the usual dot product in equation (\ref{svm}) by the Gaussian radial basis function (RBF)
\begin{equation}
    K_\eta(\mathbf{v}_i,\mathbf{v}_j)=\text{exp}(-\eta||\mathbf{v}_i-\mathbf{v}_j||^2_2)
\label{rbf}
\end{equation}
for any vectors $\mathbf{v}_i,\mathbf{v}_j$, where $\eta>0$ is a global parameter. Readers are referred to \cite{svm} for more details of the SVM.

Once the minimization problem (\ref{svm}) is solved, the QC-SPHARM classification machine is ready. For any new unlabelled subject $x$, the Hipp surface $\hat{M_x}$ is segmented and then registered to the template mean surface $M_0$ by imposing the SPHARM coefficients onto $S_0$ as in equation (\ref{mesh_reg}). The registered surface $M_x$ is thus available and hence the feature vector $\mathbf{c}_x$, is computed as defined in equation (\ref{featurevector_composed}). The discriminating feature vector $\widetilde{\mathbf{c}}_x$ is then extracted by the t-test as described above and finally, the class label for subject $x$ is given by
\begin{equation}
    y^x = \text{sign}(K_\eta(\mathbf{w},\widetilde{\mathbf{c}}_x)+b)
\label{labelling}
\end{equation}
where $\text{sign}(\cdot)$ is the sign function given by
\begin{equation}
    \text{sign}(x)=\begin{cases}1,&\text{ if }x\geq0,\\-1,&\text{ if }x<0.\end{cases}
\end{equation}

\section{Model Evaluation}

In this section, the classification results on the two datasets (ref. to (\ref{tab:database})) are investigated. For the dataset A, we randomly separate the data into a training group and a testing group. For each class, the training group has $70$ subjects and the testing group contains the remaining $40$ subjects. Combining the two different classes, we have altogether $140$ subjects in the training group and $80$ subjects in the testing group. Similar separation is done to the dataset B such that the training group contains $15$ subjects from each class and the testing group contains the remaining $5$ subjects from each class. The training group is used to build the classification machine and the testing group is used to evaluate the accuracy of the machine. The random process is repeated for $1,000$ times to reduce bias to any particular data separation. In all the following discussions, the accuracy of the model is taken to be the mean classification accuracy over the $1,000$ iterations.

There are several parameters in the proposed model. In particular, we set $\eta=1$ in the RBF (\ref{rbf}) for the SVM, and $(\alpha,\beta,\gamma)=(0.1,0.1,1)$ for the shape index weighting (\ref{shape_index}). The SPHARM coefficients for each Hipp are restricted up to order $l=31$. The threshold parameter $p_{cut}$ is varied along the range $[0.0001,0.5]$. The variation of the accuracy with varying parameters is studied. The results are also compared with other classification criterion, including the change in volume and the SPHARM coefficients, instead of the proposed shape index. 

\subsection{The classification accuracy}

\begin{figure}
    \centering
    \includegraphics[width=.9\textwidth]{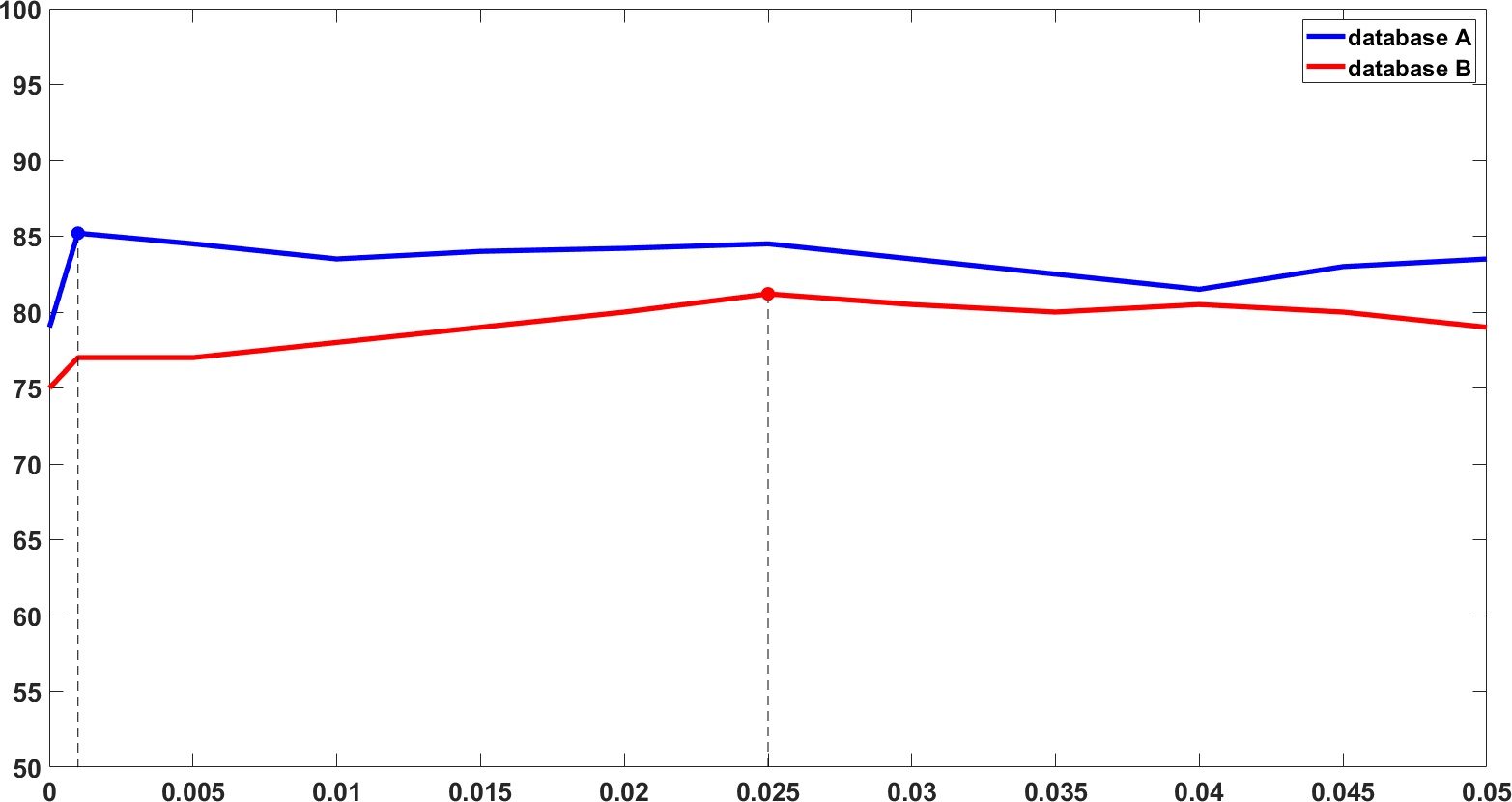}
    \caption{Variation in the classification accuracy ($\%$) versus the change in threshold parameter $p_{cut}$}
    \label{fig:vary_p}
\end{figure}

For the dataset A, the best classification accuracy is obtained at the threshold parameter $p_{cut}=0.001$, giving a correct classification rate of $85.2\%$, with a sensitivity (i.e. correctly detecting AD) of $86.5\%$ and a specificity (correctly detecting NC) of $83.9\%$. For the dataset B, the best accuracy is $81.2\%$ obtained at $p_{cut}=0.025$, with sensitivity $80.5\%$ (correctly detecting aMCI-AD) and specificity $81.9\%$ (correctly detecting aMCI-stable). Table (\ref{tab:compose_feature}) records the details of the classification accuracy and figure (\ref{fig:vary_p}) plots the accuracy of the model versus the variation of the thresholding $p_{cut}$.

\begin{table}[h]
\makebox[1 \textwidth][c]{\resizebox{1.05 \textwidth}{!}{
    \centering
    \begin{tabular}{|c|c|c|c|c|c|c|c|}
    \hline
    \multirow{2}{*}{Database} & \multirow{2}{*}{$p_{cut}$} & \multicolumn{3}{|c|}{$\#$ of features incorporated} & \multirow{2}{*}{Sensitivity} & \multirow{2}{*}{Specificity} & \multirow{2}{*}{Accuracy}  \\
    \cline{3-5}
    & & shape index & SPHARM coef & volume & & & \\
    \hline
    A & 0.001 & 4,447 & 50 & 1 & 86.5$\%$ & 83.9$\%$ & 85.2$\%$ \\
    \hline
    B & 0.025 & 337 & 425 & 0 & 80.5$\%$ & 81.9$\%$ & 81.2$\%$ \\
    \hline
    \end{tabular}}}
    \caption{Classification accuracy of the proposed model incorporating with other features including the SPHARM coefficients of the Hipp surface and the Hipp volume change}
    \label{tab:compose_feature}
\end{table}

\subsection{Comparison with other algorithms}

To better validate the classification power of the QC-SPHARM model, we compare our proposed algorithm with other approaches for early AD diagnosis, in particular, the QC model proposed in \cite{qcpaper_1}, Hipp-volume-based method and the SPHARM method. 

In the QC model proposed in \cite{qcpaper_1}, the longitudinal data (i.e. different time frame) of each subject is obtained and the surface correspondences are constructed using the center-line algorithm. The shape index is used as the discriminating features on the surfaces and a t-test is applied to extract the vertices having high discriminating power. Lastly, a $L^2$ norm based metric is defined on the feature vector space to set up a discriminating criterion to classify each subject. Readers are referred to the paper for more details. It is noted that since there is no longitudinal data available for both database, we replace the surface correspondences elaborated in \cite{qcpaper_1} by the SPHARM correspondences as mentioned in the last sections. 

In the traditional Hipp-volume-based method, the subject is scanned multiple times in an interval of half year to one year. The Hipp volumetric data is segmented from the 3D image and the change in relative Hipp volume is recorded. Once a database of pre-labelled data is available, the mean of the Hipp volume change can be calculated for the NC class and for the AD class respectively. Labelling a new subject is therefore to check if the Hipp volume change is closer to the mean of the NC class or that of the AD class. However, as for early AD diagnosis, it is undesirable to take multiple brain scans over long time before a diagnosis result is available. Instead, we compute relative Hipp volume by dividing the Hipp volume by the total brain mask volume. The relative volume should be small if the Hipp is shrinking, and thus is believed to be helpful to the classification.

In the SPHARM method, the SPHARM coefficients of the Hipp surface of each subject is used directly as the discriminating features. In particular, we use the SPHARM coefficients up to order $l=31$, which corresponds to $961$ terms in the equation (\ref{mesh_reg}). Since each $\mathbf{r}^m_{l,i}$ is a 3-dimensional vector corresponding to the $x-,y-$ and $z-$coordinate respectively, and each term is a complex number which can be separated into the real part and the imaginary part, altogether we have $961\times 3\times 2=5,766$ features on each Hipp surface. The bagging-incorporated t-test is applied on the features to extract those with high discriminative power. Then the SVM (\ref{svm}) with the Gaussian RBF kernel (\ref{rbf}) is applied to build the classification machine. 

The above three models are applied to both the dataset A and the dataset B and the results are compared with that of the QC-SPHARM model. The results are recorded in table (\ref{tab:comparison}).

\begin{table}[h]
\makebox[1 \textwidth][c]{\resizebox{1.05 \textwidth}{!}{
    \centering
    \begin{tabular}{|c|c|c|c|c|c|c|}
    \hline
    Database & $\#$ training & $\#$ testing & Method & Sensitivity & Specificity & Accuracy \\
    \hline
    \multirow{4}{*}{A} & \multirow{4}{*}{140} & \multirow{4}{*}{80} & QC-SPHARM & {\bf 86.5$\%$} & {\bf 83.9$\%$} & {\bf 85.2$\%$} \\
    \cline{4-7}
    & & & QC & 82.0$\%$ & 79.5$\%$ & 80.8$\%$ \\
    \cline{4-7}
    & & & Volume & 71.6$\%$ & 75.2$\%$ & 73.4$\%$ \\
    \cline{4-7}
    & & & SPHARM & 78.6$\%$ & 76.1$\%$ & 77.4$\%$ \\
    \hline
    \multirow{3}{*}{B} & \multirow{3}{*}{30} & \multirow{3}{*}{10} & QC-SPHARM & {\bf 80.5$\%$} & {\bf 81.9$\%$} & {\bf 81.2$\%$} \\
    \cline{4-7}
    & & & QC & 77.2$\%$ & 78.4$\%$ & 77.8$\%$ \\
    \cline{4-7}
    & & & Volume & 66.3$\%$ & 65.5$\%$ & 65.9$\%$ \\
    \cline{4-7}
    & & & SPHARM & 75.3$\%$ & 75.2$\%$ & 75.1$\%$ \\
    \hline
    \end{tabular}}}
    \caption{Comparison of the classification accuracy among the proposed model, the volume-based model and the SPHARM model on the dataset A and the dataset B}
    \label{tab:comparison}
\end{table}

From the table (\ref{tab:comparison}), our proposed QC-SPHARM model consistently outperforms the other three methods. In classifying the dataset A, our algorithm has a 5-12$\%$ advantage in accuracy over other methods. This validates that combining different type of geometric distortions are more discriminating than a single-type measurement. In classifying the dataset B, our algorithm has a 4-16$\%$ advantage over the other methods. In particular, we observe that the difference in Hipp volume between the aMCI-stable class and the aMCI-AD class is just mild, compared to that in geometric distortions. After all, it is now evident that the proposed geometric distortions are more effective in discriminating between (possible) AD patients and NC/aMCI-stable subjects.

\subsection{Visualizing the shape index}

\begin{figure}
    \centering
    \begin{subfigure}{.45\textwidth}
        \centering
        \includegraphics[width=.9\textwidth]{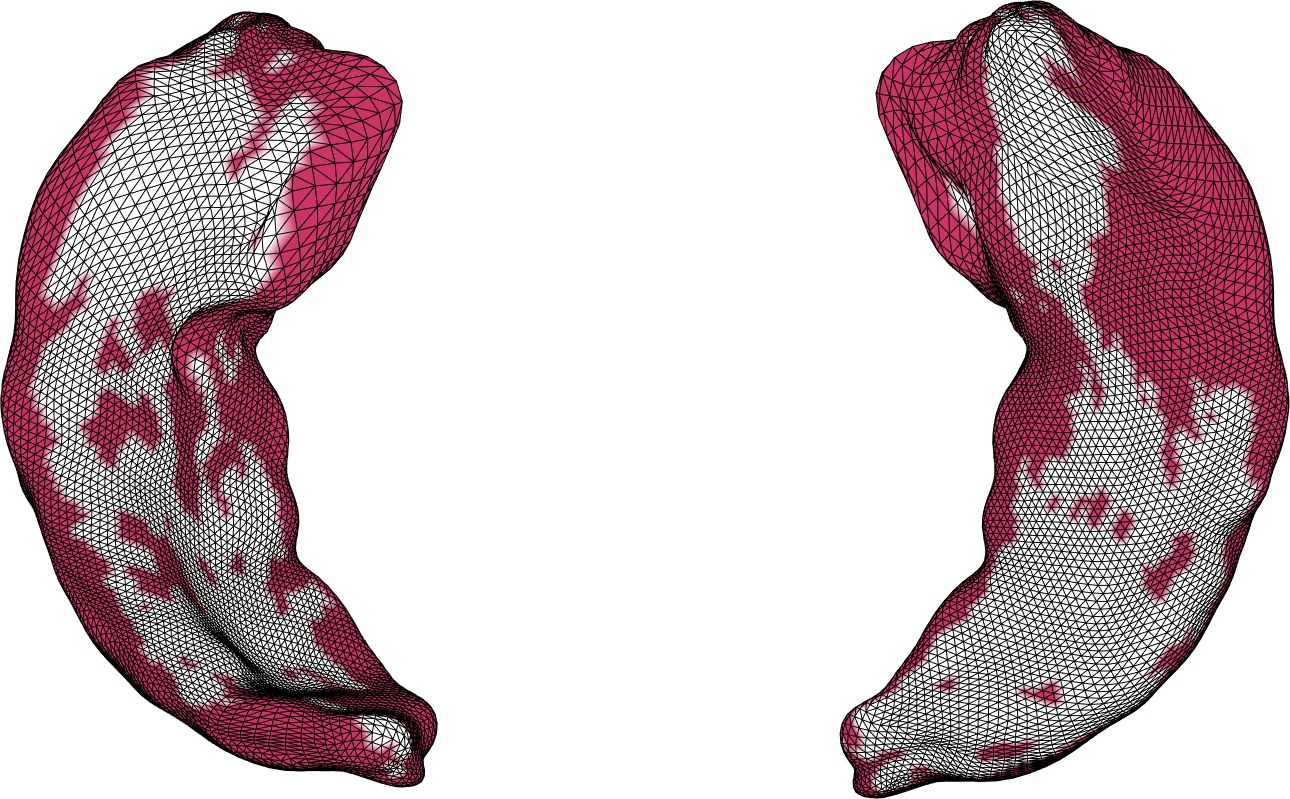}
        \caption{dataset A}
    \end{subfigure}
    \begin{subfigure}{.45\textwidth}
        \centering
        \includegraphics[width=.9\textwidth]{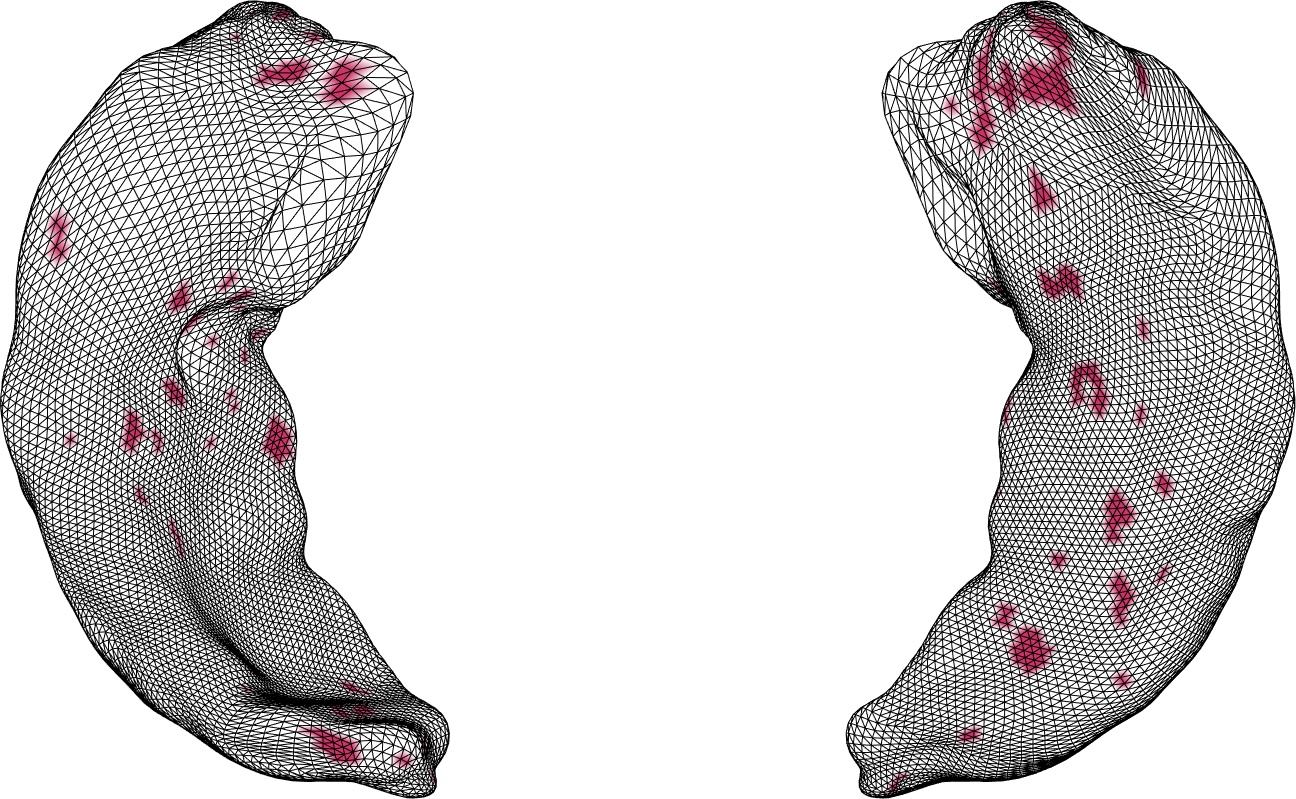}
        \caption{dataset B}
    \end{subfigure}
    \caption{Highlight of the statistical significant region $\Omega$ (in red) on a sample Hipp surface}
    \label{fig:omega}
\end{figure}

The shape index $E_{shape}$ (in the equation (\ref{shape_index}) included in the QC-SPHARM model can be visualized on the Hipp surface. In particular, the statistical significant region $\Omega$, restricted on the shape index space $E_{shape}$, can be highlighted on the Hipp surface. However, note that even by setting a constant threshold parameter $p_{cut}$ in each of the $1,000$ iterations of random separation of the database, the composition of $\Omega$ is still different due to the variation of the training dataset. Therefore, we randomly pick one result from the $1,000$ iterations. For the dataset A, the threshold parameter $p_{cut}=0.001$ corresponds to $4,447$ vertices being included in the statistical significant region $\Omega$. For the dataset B, $p_{cut}=0.025$ corresponds to $377$ vertices being included in $\Omega$. The region $\Omega$ is highlighted on a sample Hipp surface in the figure (\ref{fig:omega}).

From the above results, it can be seen that the two classes in each database show significantly different local geometric distortion patterns from the mean template surface. The difference between NC and AD subjects is larger than that between aMCI-stable subjects and aMCI-AD subjects, as a larger region is recorded to achieve a higher discrimative power for the former case. As the two classes in the dataset B are both in the stage of aMCI, such result is expected indeed. In classifying between AD and NC subjects, most significant difference in the local geometric distortion patterns happen on both tips of the Hipp surface, as well as the curvy regions in the middle part of it. In the classification between aMCI-stable subjects and aMCI-AD subjects, it is observed that about half (186 out of 377) of those selected vertices in $\Omega$ are also included in $\Omega$ while classifying between AD and NC subjects. We believe this finding is helpful for further clinical research.

\section{Conclusion}

This work presents a new model, the QC-SPHARM, using the hippocampus (Hipp) surface data for early diagnosis of the Alzheimer's Disease (AD). The study regards both the classification between AD patients and normal control (NC) subjects, as well as between amnestic mild cognitive impairment (aMCI) patients who would probably advance into AD in a two-year period and who would rather remain stable. The model represents the Hipp surface of each subject using the spherical harmonics (SPHARM) theory. A template surface of a normal Hipp is created from the database of NC subjects using the SHREC scheme. Each Hipp surface in the database is then registered to the template surface. Afterwards, geometric distortions including the conformality distortions, the curvatures distortions, the SPHARM coefficients and the volume distortion of each Hipp surface from the template surface are computed to formulate a feature vector for each subject. After applying a bagging-strategy-incorporated t-test on the feature vectors to extract those vertices having discriminating power, the support vector machine (SVM) with the Gaussian radial basis function (RBF) kernel is utilized to build the classification machine. 

The proposed model is tested with two datasets collected from the ADNI database, one of size 220 and another one of size 40. The results show that the model can achieve 81-85$\%$ accuracy on the two database, which is consistently higher than using other models including the QC model \cite{qcpaper_1}, the classical Hipp-volume based method and the SPHARM method by a significant margin. In the future, the model can be put on a deep learning setting to further boost both the classification accuracy and the processing time.


\begin{thebibliography}{00}
\bibitem{AD_population_1}
Koedam, Esther LGE, et al. ``Early-versus late-onset Alzheimer's disease: more than age alone.'' Journal of Alzheimer's Disease 19.4 (2010): 1401-1408.
\bibitem{AD_population_2}
Calderon-Garciduenas, Lilian, et al. ``Hallmarks of Alzheimer disease are evolving relentlessly in Metropolitan Mexico City infants, children and young adults. APOE4 carriers have higher suicide risk and higher odds of reaching NFT stage V at $\leq$40 years of age.'' Environmental research 164 (2018): 475-487.
\bibitem{AD_population_3}
Evans, Denis A., et al. ``Prevalence of Alzheimer's disease in a community population of older persons: higher than previously reported.'' Jama 262.18 (1989): 2551-2556.
\bibitem{AD_population_4}
Hebert, Liesi E., et al. ``Age-specific incidence of Alzheimer's disease in a community population.'' Jama 273.17 (1995): 1354-1359.
\bibitem{AD_Hipp_1}
Jack, Clifford R., et al. ``MR‐based hippocampal volumetry in the diagnosis of Alzheimer's disease.'' Neurology 42.1 (1992): 183-183.
\bibitem{AD_Hipp_2}
Jin, Kunlin, et al. ``Increased hippocampal neurogenesis in Alzheimer's disease.'' Proceedings of the National Academy of Sciences 101.1 (2004): 343-347.
\bibitem{AD_Hipp_3}
Hyman, Bradley T., et al. ``Alzheimer's disease: cell-specific pathology isolates the hippocampal formation.'' Science 225.4667 (1984): 1168-1170.
\bibitem{Hipp_volume_1}
Colliot, Olivier, et al. ``Discrimination between Alzheimer disease, mild cognitive impairment, and normal aging by using automated segmentation of the hippocampus.'' Radiology 248.1 (2008): 194-201.
\bibitem{Hipp_volume_2}
Chupin, Marie, et al. ``Fully automatic hippocampus segmentation and classification in Alzheimer's disease and mild cognitive impairment applied on data from ADNI.'' Hippocampus 19.6 (2009): 579-587.
\bibitem{Hipp_volume_3}
Laakso, M. P., et al. ``Hippocampal volumes in Alzheimer's disease, Parkinson's disease with and without dementia, and in vascular dementia: An MRI study.'' Neurology 46.3 (1996): 678-681.
\bibitem{Hipp_volume_4}
Convit, A., et al. ``Specific hippocampal volume reductions in individuals at risk for Alzheimer’s disease.'' Neurobiology of aging 18.2 (1997): 131-138.
\bibitem{Hipp_volume_5}
Bobinski, Matthew, et al. ``The histological validation of post mortem magnetic resonance imaging-determined hippocampal volume in Alzheimer's disease.'' Neuroscience 95.3 (1999): 721-725.
\bibitem{Hipp_volume_6}
Schuff, N., et al. ``MRI of hippocampal volume loss in early Alzheimer's disease in relation to ApoE genotype and biomarkers.'' Brain 132.4 (2009): 1067-1077.
\bibitem{AD_Hipp_compare}
Cuingnet, Remi, et al. ``Automatic classification of patients with Alzheimer's disease from structural MRI: a comparison of ten methods using the ADNI database.'' neuroimage 56.2 (2011): 766-781.
\bibitem{AD_CNN}
Li, S., et al. ``Hippocampal shape analysis of Alzheimer disease based on machine learning methods.'' American Journal of Neuroradiology 28.7 (2007): 1339-1345.
\bibitem{Hipp_landmark}
Lui, Lok Ming, et al. ``Shape-based diffeomorphic registration on hippocampal surfaces using beltrami holomorphic flow.'' International Conference on Medical Image Computing and Computer-Assisted Intervention. Springer, Berlin, Heidelberg, 2010.
\bibitem{spharm}
Kazhdan, Michael, Thomas Funkhouser, and Szymon Rusinkiewicz. ``Rotation invariant spherical harmonic representation of 3 d shape descriptors.'' Symposium on geometry processing. Vol. 6. 2003.
\bibitem{Hipp_spharm_1}
Gutman, Boris, et al. ``Disease classification with hippocampal shape invariants.'' Hippocampus 19.6 (2009): 572-578.
\bibitem{Hipp_spharm_2}
Gerardin, Emilie, et al. ``Multidimensional classification of hippocampal shape features discriminates Alzheimer's disease and mild cognitive impairment from normal aging.'' Neuroimage 47.4 (2009): 1476-1486.
\bibitem{Hipp_spharm_3}
Gutman, Boris, et al. ``Hippocampal surface analysis using spherical harmonic function applied to surface conformal mapping.'' 18th International Conference on Pattern Recognition (ICPR'06). Vol. 3. IEEE, 2006.
\bibitem{Hipp_spharm_4}
Shen, Li, et al. ``Morphometric MRI study of hippocampal shape in MCI using spherical harmonics.'' Alzheimer's $\&$ Dementia: The Journal of the Alzheimer's Association 1.1 (2005): S47.
\bibitem{AccuBrain}
Abrigo J, et al. Standardization of hippocampus volumetry using automated brain structure volumetry tool for an initial alzheimer's disease imaging biomarker. Acta Radiologica. 2019; 60(6):769-776 .
\bibitem{itk-snap}
Paul A. Yushkevich, Joseph Piven, Heather Cody Hazlett, Rachel Gimpel Smith, Sean Ho, James C. Gee, and Guido Gerig. ``User-guided 3D active contour segmentation of anatomical structures: Significantly improved efficiency and reliability.'' Neuroimage 2006 Jul 1;31(3):1116-28
\bibitem{SHREC}
Shen, Li, et al. ``Efficient registration of 3D SPHARM surfaces.'' Fourth Canadian Conference on Computer and Robot Vision (CRV'07). IEEE, 2007.
\bibitem{qcpaper_1}
Chan, Hei Long, Hangfan Li, and Lok Ming Lui. ``Quasi-conformal statistical shape analysis of Hipp surfaces for Alzheimer's disease analysis.'' Neurocomputing 175 (2016): 177-187.
\bibitem{spharm_morphing}
Shen, Li, Hany Farid, and Mark A. McPeek. ``Modeling three‐dimensional morphological structures using spherical harmonics.'' Evolution: International Journal of Organic Evolution 63.4 (2009): 1003-1016.
\bibitem{qcpaper_2}
Chan, Hei Long, and Lok Ming Lui. ``Detection of n-dimensional shape deformities using n-dimensional quasi-conformal maps.'' J. Geom. Imaging Comput 1.4 (2014): 395-415.
\bibitem{bagging}
Breiman, Leo. ``Bagging predictors.'' Machine learning 24.2 (1996): 123-140.
\bibitem{svm}
Cristianini, Nello, and John Shawe-Taylor. ``An introduction to support vector machines and other kernel-based learning methods.'' Cambridge university press, 2000.

\end{thebibliography}
\end{document}